\documentclass{article}
\usepackage{ijcai19}
\usepackage[english]{babel}

\usepackage{times}
\usepackage{soul}
\usepackage{url}
\usepackage[utf8]{inputenc}
\usepackage[small]{caption}
\usepackage{graphicx}
\usepackage{amsmath}
\usepackage{booktabs}
\usepackage{graphicx}
\usepackage{hyperref}
\title{Comparative Evaluation of 3D and 2D Deep Learning Techniques for Semantic Segmentation in CT Scans}

\author{
Abhishek Shivdeo\and
Rohit Lokwani\and
Viraj Kulkarni\and
Amit Kharat\and Aniruddha Pant
\affiliations
DeepTek Inc\\
\emails
}

\begin{document}

\maketitle

\begin{abstract}
Image segmentation plays a pivotal role in several medical-imaging applications by assisting the segmentation of the regions of interest. Deep learning-based approaches have been widely adopted for semantic segmentation of medical data. In recent years, in addition to 2D deep learning architectures, 3D architectures have been employed as the predictive algorithms for 3D medical image data. In this paper, we propose a 3D stack-based deep learning technique for segmenting manifestations of consolidation and ground-glass opacities in 3D Computed Tomography (CT) scans. We also present a comparison based on the segmentation results, the contextual information retained, and the inference time between this 3D technique and a traditional 2D deep learning technique. We also define the area-plot, which represents the peculiar pattern observed in the slice-wise areas of the pathology regions predicted by these deep learning models. In our exhaustive evaluation, 3D technique performs better than the 2D technique for the segmentation of CT scans. We get dice scores of 79\% and 73\% for the 3D and the 2D techniques respectively. The 3D technique results in a 5X reduction in the inference time compared to the 2D technique. Results also show that the area-plots predicted by the 3D model are more similar to the ground truth than those predicted by the 2D model. We also show how increasing the amount of contextual information retained during the training can improve the 3D model’s performance.

\end{abstract}

\section{Introduction}
Medical imaging techniques like X-rays, Magnetic Resonance Imaging (MRI), Computed Tomography (CT), etc., provide precise anatomy of a human body and thus help detect abnormalities present in the body \cite{umar2019review}. Effective and early identification of regions of infection in medical images can play a crucial role in assisting the doctors for the treatment of various pathologies. For instance, observing X-rays and finding early signs of pneumonia, which causes around 50,000 deaths per year in the US \cite{centers2012pneumonia}, and treating it in time can save many lives. However, X-rays compress a 3D volume into a single 2D image, which causes loss of information. X-rays also lack specificity when pathology regions are concealed by overlapping tissues, bones, or bad contrast environments when detecting pathologies like COVID-19 \cite{zhang2020viral}. Thus, this makes understanding and identifying a pathology using high-resolution CT scans a sought after medical diagnosis technique \cite{doi2007computer}. 

CT scans are 3D medical images that comprise several slices or images, similar to X-rays, stacked upon each other, which combine to give us a volumetric representation of the interior aspects of our body \cite{karatas2014three}. Nevertheless, classifying and marking regions of interest in CT scans needs significant effort from the radiologists. Hence, automated detection and segmentation of pathologies in CT scans, to reduce a radiologist's involvement, is seen as an essential tool for diagnosing and treating a disease. 

The rapid research and development in machine learning, graphics processing technologies and the availability of large amounts of data \cite{rodriguez2016general} \cite{minsky1961steps} \cite{cockburn2018impact} \cite{yang2018research} have improved the field of Computer vision \cite{lu2020survey}  [6]. The availability of high-quality medical image datasets \cite{kohli2017medical} combined with rapid advancement in CNN based architectures has led to an increase in the adoption of deep learning models to assist radiologists in evaluating CT scans. Deep learning models are used to detect, classify, and segment fractures, tumors, and other pathologies in CT scans. However, there is a lot of variation in the contrast of images based on different radiation doses \cite{smith2019international} \cite{trattner2014standardization} given to patients. The quality of CT scanners in different hospitals, and slice thickness can also differ from scan to scan in a multi-sourced dataset, making it challenging to train machine learning models for CT scans. Also, selecting a proper CT window, by manipulating the Hounsfield unit (HU) values in a CT, can affect the model’s performance \cite{xue2012window}. A deep learning model needs to be robust enough to handle these variations or it may experience a covariance shift when it is tested on out-of-source data. Studies have tried to mitigate these effects by trying out image noise reduction methods to reduce the radiation dose of CT imaging \cite{willemink2019evolution} \cite{yang2018low}. 

Researchers have adopted both 2D as well as 3D approaches \cite{haque2020deep}. Studies conducted by Zhou et al. \cite{zhou2018performance} compare the segmentation performance of 2D and 3D deep learning-based approaches using conventional segmentation metric such as dice score. In a 2D deep learning technique \cite{weston2019automated}, the input to the model is a single 2D image, whereas, in a 3D deep learning technique \cite{zhao2020mss}, the model takes a 3D volume as its input. Both these approaches employ a Fully Connected Network (FCN) for segmentation. As opposed to training slice by slice in 2D FCNs, the 3D FCNs models analyze volumetric input data and utilize the global features in between the CT slices. Just like how a word in a sentence gives a clue about what the next word could be \cite{akbik2018contextual} \cite{hassan2017deep}, a CT scan’s slice can give a clue about the shape of the pathology in its adjacent slices. This is because, in most of the pathologies, (consolidation and ground-glass-opacities here) the regions of interest (ROI) or the areas of the pathology in a scan follow a continuous pattern. As the CT scans are captured in a particular order, it is observed that continuity of manifestations exists in the adjacent slices, we can observe the same in Figure 6. Using 3D models for the segmentation of CT scans is similar to using LSTMs with the attention module \cite{olah2016attention} for the formation of a sentence. This contextual information is lost when we use 2D CT scans because the 2D model predicts the outcome by considering an individual slice as a single data point, thus their prediction is not affected by the adjacent slices, which is evident from the results of our experiments (Figure 7).

In our paper, we propose a 3D technique for the segmentation of consolidation and ground-glass opacities in CT scans, and also present a comparison between this 3D technique and a traditional 2D technique. We compare the dice scores between the predicted masks and the radiologists’ annotations for both these techniques. We also plot the areas of the predicted masks versus the position of the slice in the CT scan for both 2D and 3D techniques, and compare it with the ground truth area-plots. We also compare the inference time for these two techniques, which is an important factor when a model runs inference in real-life situations post-deployment.

\section{Related Work}
Recent studies have emphasized the use of deep learning for medical imaging analysis. The problems solved using deep learning can be broadly classified into image classification and semantic segmentation. Convolutional Neural Networks (CNNs) are commonly used for image classification. Badea et al. \cite{badea2016use} used LeNet \cite{lecun1998gradient} and NiN (Network in Network) \cite{lin2013network} for classifying burns on the human body from images of size 320 x 240 captured using a camera and achieved an accuracy of 75.91\% and 58.01\% for classification of Skin vs. Burn and Skin vs. Light Burn vs. Serious Burn, respectively. Polsinelli et al. \cite{polsinelli2020light} used SqueezeNet, a CNN architecture, for classifying CT scan’s slices into COVID-19 or non-COVID-19 with an accuracy of 85\%. The classification process is simpler than segmentation because in classification all the pixels in a single image need to be grouped into a single class. While in semantic segmentation, each pixel needs to be assigned a class.

Image segmentation was initially solved using conventional image processing approaches. Okada et al.\cite{okada2015abdominal} proposed an image processing technique for multi-organ segmentation in which he used statistical shape modeling and probabilistic atlas and the segmentation of organs was done by combining the intra-organ information with the inter-organ correlation to get an average dice coefficients of 92\% for the liver, spleen, and kidneys, and a dice coefficient of 73\% and 67\% for the pancreas and gallbladder, respectively. This conventional approach demonstrated highly accurate multiple organ segmentation techniques for CT scans and presented a detailed evaluation of the observations. 

After the development of encoder-decoder \cite{badrinarayanan2017segnet} architectures, they were commonly used for segmentation purposes. U-Net, a 2D deep learning approach, proposed by Ronneberger et al. \cite{ronneberger2015u}, segmented a single (512, 512) image in under a second on an NVidia Titan GPU. U-Net was fast, efficient, and accurate and thus was the first widely used deep learning architecture for image segmentation tasks for medical image data \cite{ronneberger2015u}. Christ et al. \cite{christ2017automatic} cascaded two FCNs to segment out the liver and its lesions from CT and MRI scans and achieved an accuracy of around 94\% on the validation set in under 100 seconds per volume. Almotairi et al. \cite{almotairi2020liver} proposed another deep learning architecture, SegNet, that employed a trained VGG-16 image classification network as its encoder, and had a corresponding decoder architecture for pixel-wise classification at the end, which was able to achieve an accuracy of 99.99\% for segmenting a liver tumor. For these 2D  approaches, slices of the MRI and CT scans present in the dataset were treated as individual 2D images, which means that the 3D volumetric data is transformed into a 2D planar data. Other such 2D based image segmentation approaches were implemented \cite{zhou2016first} \cite{long2015fully} \cite{de2015deep} \cite{roth2015deep} \cite{cha2016urinary}.

Zhou et al. \cite{zhou2016three} proposed a segmentation approach in which the 2D slice-wise results were later combined using 3D majority voting, where a simple encoder-decoder network was combined to be a part of an all-in-one network which could segment out complicated multiple organs; it correctly segmented 89\% of the voxels from the CT scans.

However, in recent years, due to improved 3D convolution architectures and advancements in computational power (GPUs), training of highly complex	3D deep learning models having a 3D volume as its input, has become much more accurate, efficient, and faster \cite{kayid2018performance}. Cicek et al. \cite{cciccek20163d} proposed a 3D U-Net architecture that predicted volumetric segmentation using 2D annotated slices. The average Intersection over Union (IoU) achieved was 0.863. They \cite{cciccek20163d} were able to annotate unseen data as well as densify the sparsely annotated data. 

Milletari et al. \cite{milletari2016v} proposed V-Net: a novel fully convolutional neural network for volumetric medical image segmentation which gave an average dice score of 86\% to segment out the prostate depicted in 30 MRI scans. These MRIs were converted into a constant volume of  128 × 128 × 64 using B-spline interpolation, which alters the global features during the conversion and can have detrimental effects on the training. 

VoxResNet proposed by Chen et al. \cite{chen2016voxresnet}, which borrows the spirit of deep residual learning in 2D image recognition tasks, and is extended into a 3D variant for handling volumetric data, has also been successfully applied for 3D medical image segmentation tasks. Allan et al. \cite{alalwan60efficient} proposed another 3D FCN model architecture called “3D-DenseUNet-569” for liver and tumor segmentation which used Depthwise Separable Convolution (DS-Conv) as opposed to traditional convolution.

Although between 2D and 3D, 3D FCN provides us with more accurate results, it is more complex and requires higher memory along with greater computational resources \cite{li2018h}. The higher complexity restrains the model from training a larger dataset efficiently. Moreover, the high memory footprint leads to reduced network depth and filter size, which adversely affects the performance of the model \cite{simonyan2014very}.

\section{Data}
For our experiments, we obtained 182 CT scans from 2 private Indian hospitals. These CT scans have non-uniform volumes and are annotated for consolidation and ground-glass-opacities \cite{chung2020ct}. Our team of expert radiologists marked out the regions of infections, in the form of free-hand annotations, which served as the ground truth for our model. These precise annotations were done using the ITK-snap tool, an open-source free-hand annotation tool \cite{yushkevich2016itk}. Some examples of the CT slices and their superimposed masks are showcased in the Figure \ref{fig: Left: CT Slice, Right: Free hand annotated CT Slice} (a) and Figure 1 (b).

\begin{figure}[hbt!]
\begin{center}
\includegraphics[scale=0.4]{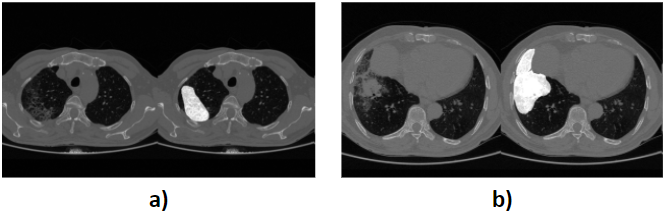}
\caption{(a,b): CT Slice (Left), Free hand annotated CT Slice (Right)}
\label{fig: Left: CT Slice, Right: Free hand annotated CT Slice}
\end{center}
\end{figure}

\begin{table}[hbt!]
\begin{center}
\begin{tabular}{ | c | c | c |}
\hline
Dataset & Number of CT Scans & Number of Slices\\
\hline
Training & 126 & 56387\\
Validation & 20 & 9992\\
Test  & 36 & 14727\\

\hline
\end{tabular}
\caption{Scan-level and Slice-level Dataset splits}\label{fig: Scan-level and Slice-level Dataset splits}
\end{center}
\end{table}

The positive class COVID-19 comprised consolidation and ground-glass opacities. These chest CT scans were divided into train, validation, and test datasets whose splits are given in Table 1.

\begin{figure}[hbt!]
\begin{center}
\includegraphics[scale=0.8]{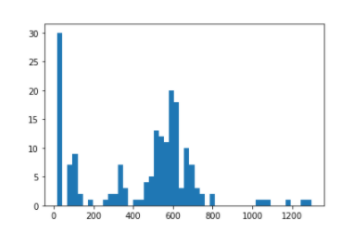}
\caption{Volume variation in the CT scans for the whole dataset}\label{fig:Volume variation in the CT scans for the whole dataset}
\end{center}
\end{figure}

The prevalence for all the datasets is 20\%, which means that there are 20\% positive slices in the total dataset. We resized all the images to a standard image size of (512, 512). Windowing, also known as grey-level mapping is the pre-processing of CT scans in which the grey-scale component of the CT slice is changed to highlight some particular features. We applied windowing to our scans \cite{lee2018practical}, for which we used the information stored in the metadata from the DICOM files of the CT scans' slices. The masks were stored as binary images of size (512, 512).
The distribution of the number of slices in these CT scans for the whole dataset can be visualized from the histogram in Figure \ref{fig:Volume variation in the CT scans for the whole dataset}

\section{Methodology}
In this paper, we follow two approaches to address image segmentation of these CT scans:

\subsection{2D Approach}

\begin{figure}[hbt!]
\includegraphics[width=\linewidth]{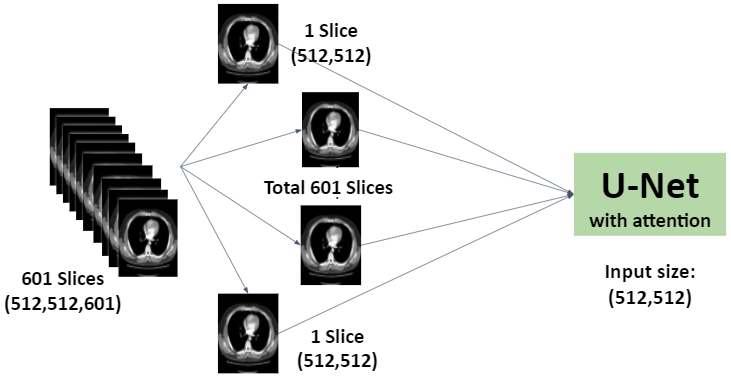}
\caption{2D Technique}
\end{figure}

In this approach, we divide the CT scan volume into separate 2D slices (Figure 3). For example, a CT scan of volume (512, 512, 601) containing 601 slices was divided into 601 individual images of (512, 512). We used the U-net with the Convolutional Block Attention Module (CBAM) \cite{woo2018cbam} to focus its attention on a region of the image and then segment the pathology from that region of the image. U-Nets are fully convolutional networks having skip connections between encoder and decoder which provide deconvolution layers with important features \cite{hesamian2019deep}. We used Xception \cite{chollet2017xception} as the encoder having depthwise separable convolutions and residual connections. The model was trained using ADAM as its optimizer with an initial learning rate of 1e-3 and having a learning rate scheduler which reduced the learning rate to 1/3 the original learning rate, every 5 epochs. We used dice loss as the loss function. Our architecture had 38 million trainable parameters.

\subsection{3D Approach}
In the 3D approach, instead of assigning 2D images as the input for the model, we provide 3D volumes as to input for the model, and the corresponding stacked annotated slices as the label. For our 3D approach, we use V-Net \cite{milletari2016v}, which is a 3D implementation of U-Net. For our input size of (512, 512, 32); V-Net had 206 million trainable parameters. We used the dice loss as the loss function while training. The model was trained using ADAM as its optimizer with an initial learning rate of 1e-3 and having a learning rate scheduler which reduced the learning rate to 1/3 the original learning rate every 5 epochs. The input to V-Net is volumetric data of dimension (x, y, z) = (512, 512, 32) where x is slice's height, y is slice's width, and z is the number of slices in the depth of the volume. To prepare the input data for training V-Net, we normalize the volume of the CT scans to the same value, having a dimension equal to the input dimension of the model. To satisfy this condition without losing any information, we split CTs into smaller volumes that match the input dimensions. We can see the CT scan being split into multiple stacks of slices in Figure 4.

\begin{figure}[hbt!]
\includegraphics[width=\linewidth]{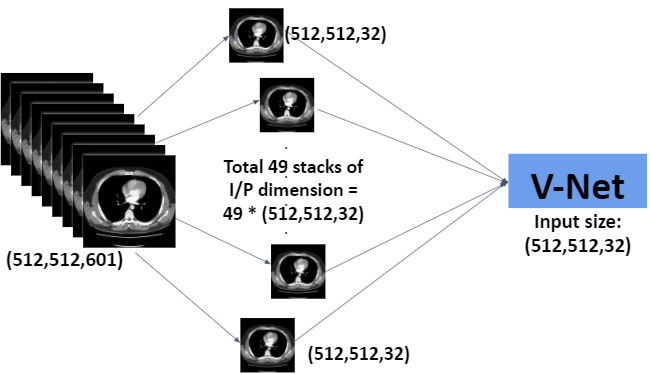}
\caption{3D Technique}
\end{figure}

There are three variables that affect the stacks creation process.\\
1. CT Volume: Number of slices in the CT scan\\
2. Stack size: Desired number of slices in the sub volume\\
3.Overlap factor = (Number of overlapping slices/stack size)
Where overlapping slices is the number of slices that are common or overlapping in adjacent stacks.

For a CT scan having 601 slices, with a stack size of 32, and the number of overlapping slices as 20 (overlap factor = 0.625), we get a list of 49 stacks having dimensions (512, 512, 32). The first stack will have indices from 0 to 32, which means that the first input datapoint for V-Net will be a CT sub volume from the 1st slice to the 32nd slice. The second input data point will be a sub volume of the CT from index 12 to 44, both inclusive. As we have 20 overlapping slices, the second stack starts from the 12th index and not the 33rd. And so on for the rest of the stacks. For the last stack, however, the indices are (576, 608), the 7 extra slices are the added paddings to keep the volume of the stack compatible with the 3D model’s input dimension. During the inference, we keep the overlap factor as 0, the list of predicted mask volumes are then stacked together. So during the inference evaluation, we have a total of 19 stacks grouped. We remove the padding when the predictions are stacked together to match the volume of the whole CT scan.

\section{Results}
We evaluate the 2D and the 3D model based on these 5 criteria.

\subsection{Dice Score}
We predicted masks for 32 scans in the test set having a prevalence of 20\% and calculated the dice score for the whole dataset. We observed the dice scores given in Table 2.

\begin{table}[hbt!]
\begin{center}
\begin{tabular}{| c | c |}
\hline
Model & Dice Score\\
\hline
2D Model & 73\%\\
3D Model & 79\%\\

\hline
\end{tabular}
\caption{Dice Scores for 2D and 3D techniques}
\end{center}
\end{table}

We took the average of the dice scores of these 32 scans. For the 2D model, we arranged the individual slices on top of each other to get the final predicted volume. For the 3D model, we combined the stacks of an individual CT and then removed the corresponding padding to match the true label volume.

\subsection{Slices Predicted masks}

Figure 5(a) and 5(b) show the predicted masks at the scan and slice level for 2D and the 3D model. We applied a threshold of 0.2 on the predictions.

\begin{figure}[hbt!]
\begin{center}
\includegraphics[scale = 0.5]{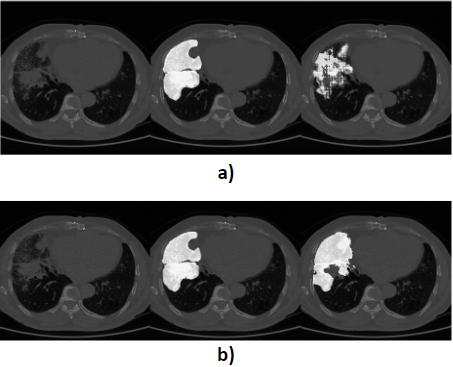}
\caption{a)3D Model, b) 2D Model\\ Left=Original CT Slice, Middle = True label superimposed, Right = Predicted Mask superimposed}
\end{center}
\end{figure}

\subsection{Area-plots}
Here, we find the area of the annotated region with respect to the total area of the image. We normalize this list of area-ratios by dividing all the values by the maximum value in the list. Next, we plot these normalized values according to the slices in the CT scan to get the plot in Figure 6, these plots are called area-plots. We follow the same procedure for the predicted masks for 2D and 3D models, by plotting their normalized area-plots according to the position of the slices in the CT scan in Figure 7, 8 respectively. We have compared the predicted area-plots for both the 2D and the 3D approach to the area-plots of the true label.

\begin{figure}[hbt!]
\begin{center}
\minipage{0.32\textwidth}
  \includegraphics[width=\textwidth]{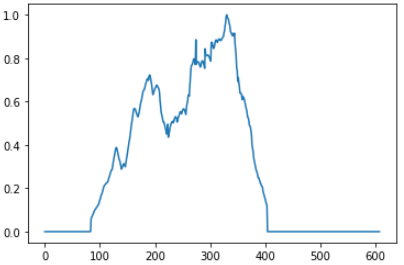}
  \caption{True Label's Area-Plot}
\endminipage\hfill
\minipage{0.32\textwidth}
  \includegraphics[width=\linewidth]{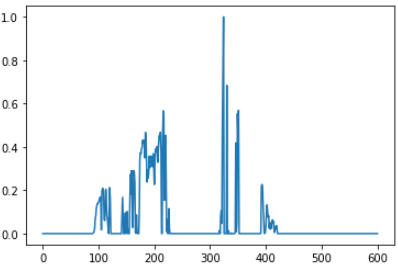}
  \caption{Area-Plot predicted by 2D Model}
\endminipage\hfill
\minipage{0.32\textwidth}%
  \includegraphics[width=\linewidth]{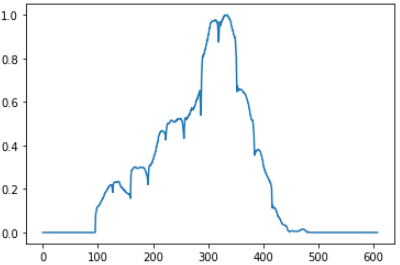}
  \caption{Area-Plot predicted by 3D Model}
\endminipage
\end{center}
% \caption{Area Plots}
\end{figure}

1. From Figure 6, we observe a continuous pattern in the manifestations of the pathology in the CT scan. Similar continuous patterns were observed for the rest of the CT scans, area-plots.\\
2. From Figure 7, we observe that there are abrupt variations in the predicted mask’s area for the slices within a CT scan when we use the 2D approach. \\
3. From Figure 8, we can see the prediction area-plot for a CT scan when using the 3D approach. We observe a smooth and continuous pattern in the masks predicted by the 3D model for the slices of that CT scan.\\

To further prove and understand point number 2, we plotted the predicted masks for consecutive slices in a CT scan using 2D and the 3D approach.

\begin{figure}[hbt!]
\includegraphics[width=\linewidth]{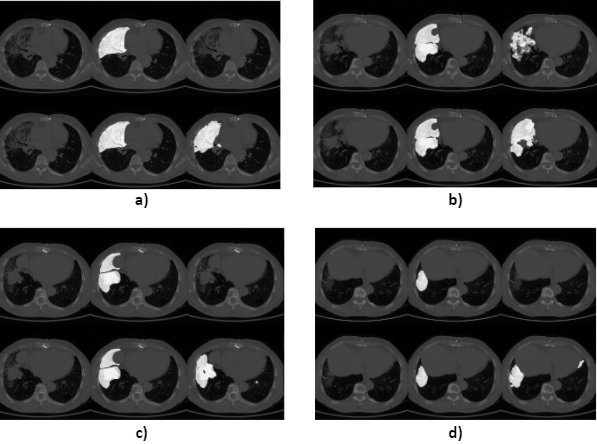}
\caption{Predicted masks in four consecutive slices (a, b, c, d).\\ Left: Original slice, Middle: True Mask, Right: Predicted Mask  \\
2D (top 3 images) and 3D (bottom 3 images) }
\end{figure}

For the four consecutive positive slices that we chose, the 3D model predicted all the slices as positive, i.e.(1,1,1,1), whereas, in the case of the 2D model, the predictions were (0,1,0,0). From Figure 9, we observe that the 3D approach predicted the masks in all 4 consecutive slices to closely match the labels. This represents the continuity in the predictions. We also observe that the shape or the area of the predicted masks do not change abruptly when we look at the adjacent slices. Although, in the masks predicted by the 2D approach, we observe that only some of the slices were marked positive. This represents the discontinuity in the predicted masks’ pattern. It indicates that the 2D approach does not consider the slices' information adjacent to the input slice, thus giving us this abrupt predicted pattern, which is independent of the slices above and below the current slice. 

\subsection{Inference time}
We calculate the inference time for 2D as well as 3D techniques for a CT scan.

\begin{table}[hbt!]
\begin{center}
\begin{tabular}{| c | c | c |}
\hline
Technique & Inference time & Inference time\\
 & With GPU &  Without GPU\\
\hline
2D & 70 & 1145\\
3D & 17 & 229\\

\hline
\end{tabular}
\caption{Inference time for 2D and 3D techniques in seconds}
\end{center}
\end{table}

Table 3 shows the inference time for a single CT scan having 709 slices with a stack size of 32 and an image size of (512, 512). We observe that there is a boost up of 5X in the 3D approach as compared to the 2D approach. For inference, we used a Tesla T4 GPU with 15GB of memory.

\subsection{Overlap Variation}
In this experiment, we changed the overlap factor to see its effect on the predictions. We kept the overlap factor as 0, 0.375, and 0.625. The predicted area-plots are given in Figure 11, 12, 13 respectively and the predicted masks are given in fig 14(a), 14(b), and 14(c) respectively. Figure 10 shows the true area-plot for the same CT scan.

\begin{figure}[hbt!]
\begin{center}
\minipage{0.32\textwidth}
  \includegraphics[width=\textwidth]{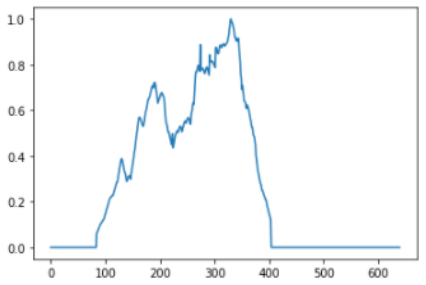}
  \caption{True Label's Area-Plot}
\endminipage\hfill
\minipage{0.32\textwidth}
  \includegraphics[width=\linewidth]{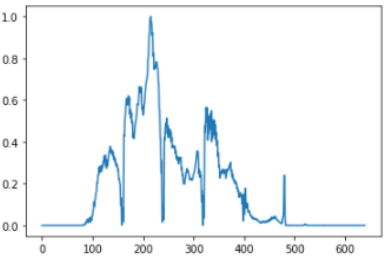}
  \caption{Area-Plot predicted by 3D Model, with overlap factor = 0}
\endminipage\hfill
\minipage{0.32\textwidth}
  \includegraphics[width=\linewidth]{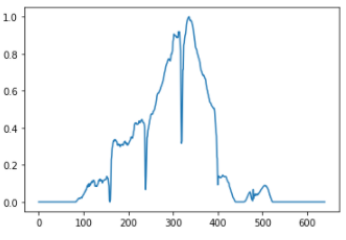}
  \caption{Area-Plot predicted by 3D Model, with overlap factor = 0.375}
\endminipage\hfill
\minipage{0.32\textwidth}%
  \includegraphics[width=\linewidth]{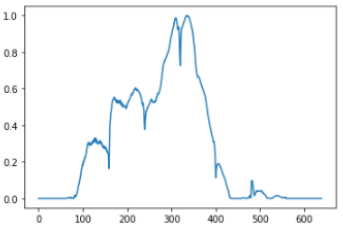}
  \caption{Area-Plot predicted by 3D Model, with overlap factor = 0.625}
\endminipage
\end{center}
% \caption{Area Plots with different overlap factors}
\end{figure}

\begin{figure}[hbt!]
\begin{center}
\includegraphics[scale=0.6]{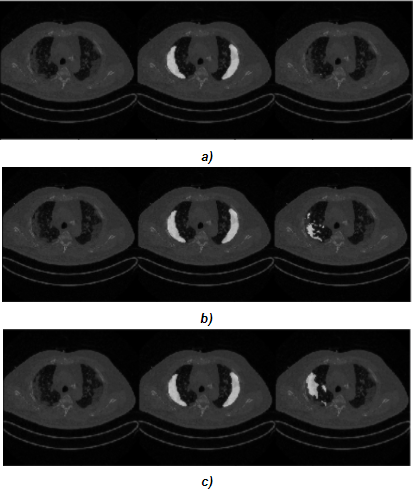}
\caption{Predicted mask for the same slice having (a) overlap factor = 0, (b)overlap factor = 0.375, (c) overlap factor = 0.625\\Left: Original slice, Middle: True Mask, Right: Predicted Mask
}
\end{center}
\end{figure}

As we increase the overlap factor, the number of overlapping slices increases. This allows the model to interpret more global features. We observe that the predicted area-plots in Figure 14 with an overlap factor 0.625 match closely to the true area-plots in Figure 10. This shift in the pattern from Figure 11 to Figure 13 demonstrates that increasing the amount of contextual information retained allows the model to perform better thus giving closer predictions to the ground truth.

\section{Conclusion}
We implement and compare two deep learning approaches for segmenting out manifestations of consolidation and ground-glass-opacities in CT scans on three major grounds: predicted masks or the segmentation results, the pattern observed in the area of the predicted masks in a CT scan, and the inference time. 

We saw that the 3D approach provided us with a better dice score than the 2D approach, thus proving to be more accurate at segmenting pathology regions. We also observed from Figure 8 that the area-plots we get using the 3D approach match closely to the original annotation area-plots against the area-plots of the 2D approach. We attribute these peculiar predicted patterns in the area-plots to the contextual information retained in the 3D stack volumes used to train the 3D model. The independent nature of the input images in the 2D model could be the reason for a discontinuous pattern in the predicted area-plots of the 2D approach. When we calculated the inference time, the 3D approach provided a boost of 5X in inference time compared to that of the 2D approach This is hugely beneficial for effective and quick diagnosis of patients in hospital settings especially the Intensive Care Units(ICUs). We conclude that this 3D stack-based approach is a better choice than the conventional 2D approach. 

Later, when we experimented with the overlap factor for the 3D approach, the predicted area-plots' patterns changed (Figure 11, 12, 13) with increasing overlap factor. However, a higher overlap factor may result in overfitting the model to the given dataset. The model gets more susceptible to a covariance-shift in the case of out-of-sample datasets when the overlap factor is high. Thus, the overlap factor should be treated as a hyperparameter for this 3D approach and should be set optimally. 

Usually, in the case of 3D segmentation for CT scans, the whole CT scan is compressed using spline interpolation \cite{milletari2016v}, or other methods, to the size of input dimension for a 3D model. This removes useful information when compressed for training and adds noise when decompressed for inference \cite{hahn2016comparison}. To counter this alteration in the original data, we split the CT into smaller stacks to retain the valuable contextual information. Thus, our batch-based 3D approach is highly adaptable to any CT volume because of its stack-based nature, where none of the information between the slices in the CT is altered. 

The segmentation results can be improved using more data, and different augmentation techniques while training. In our case study, the goal was to compare the 2D and the 3D approaches on equal grounds, and not the segmentation result itself.  In conclusion, the 3D approach that has been proposed in this paper has outperformed the traditional 2D approach in all the three criteria that we had set for evaluation.

% \begin{figure}
% \includegraphics[width=\linewidth]{Graph.png}
% \caption{Trend of predictions of Slice No. 0 to 300 from CT no. 16.}
% \end{figure}

% \printbibliography 
\bibliographystyle{ieeetr}
\bibliography{bibliography}

\begin{thebibliography}{10}

\bibitem{umar2019review}
A.~Umar and S.~Atabo, ``A review of imaging techniques in scientific
  research/clinical diagnosis,'' {\em MOJ Anat \& Physiol}, vol.~6, no.~5,
  pp.~175--183, 2019.

\bibitem{centers2012pneumonia}
C.~for Disease~Control, Prevention, {\em et~al.}, ``Pneumonia can be
  prevented--vaccines can help,'' 2012.

\bibitem{zhang2020viral}
J.~Zhang, Y.~Xie, Z.~Liao, G.~Pang, J.~Verjans, W.~Li, Z.~Sun, J.~He, Y.~Li,
  C.~Shen, {\em et~al.}, ``Viral pneumonia screening on chest x-ray images
  using confidence-aware anomaly detection,'' {\em arXiv preprint
  arXiv:2003.12338}, vol.~3, 2020.

\bibitem{doi2007computer}
K.~Doi, ``Computer-aided diagnosis in medical imaging: historical review,
  current status and future potential,'' {\em Computerized medical imaging and
  graphics}, vol.~31, no.~4-5, pp.~198--211, 2007.

\bibitem{karatas2014three}
O.~H. Karatas and E.~Toy, ``Three-dimensional imaging techniques: A literature
  review,'' {\em European journal of dentistry}, vol.~8, no.~1, p.~132, 2014.

\bibitem{rodriguez2016general}
L.~Rodr{\'\i}guez-Mazahua, C.-A. Rodr{\'\i}guez-Enr{\'\i}quez, J.~L.
  S{\'a}nchez-Cervantes, J.~Cervantes, J.~L. Garc{\'\i}a-Alcaraz, and
  G.~Alor-Hern{\'a}ndez, ``A general perspective of big data: applications,
  tools, challenges and trends,'' {\em The Journal of Supercomputing}, vol.~72,
  no.~8, pp.~3073--3113, 2016.

\bibitem{minsky1961steps}
M.~Minsky, ``Steps toward artificial intelligence,'' {\em Proceedings of the
  IRE}, vol.~49, no.~1, pp.~8--30, 1961.

\bibitem{cockburn2018impact}
I.~M. Cockburn, R.~Henderson, and S.~Stern, ``The impact of artificial
  intelligence on innovation,'' tech. rep., National bureau of economic
  research, 2018.

\bibitem{yang2018research}
L.~Yang, ``Research on application of artificial intelligence based on big data
  background in computer network technology,'' {\em Journal of Jiujiang
  Vocational \& Technical College}, vol.~392, no.~6, 2018.

\bibitem{lu2020survey}
Y.~Lu and S.~Young, ``A survey of public datasets for computer vision tasks in
  precision agriculture,'' {\em Computers and Electronics in Agriculture},
  vol.~178, p.~105760, 2020.

\bibitem{kohli2017medical}
M.~D. Kohli, R.~M. Summers, and J.~R. Geis, ``Medical image data and datasets
  in the era of machine learning—whitepaper from the 2016 c-mimi meeting
  dataset session,'' {\em Journal of digital imaging}, vol.~30, no.~4,
  pp.~392--399, 2017.

\bibitem{smith2019international}
R.~Smith-Bindman, Y.~Wang, P.~Chu, R.~Chung, A.~J. Einstein, J.~Balcombe,
  M.~Cocker, M.~Das, B.~N. Delman, M.~Flynn, {\em et~al.}, ``International
  variation in radiation dose for computed tomography examinations: prospective
  cohort study,'' {\em Bmj}, vol.~364, 2019.

\bibitem{trattner2014standardization}
S.~Trattner, G.~D. Pearson, C.~Chin, D.~D. Cody, R.~Gupta, C.~P. Hess, M.~K.
  Kalra, J.~M. Kofler~Jr, M.~S. Krishnam, and A.~J. Einstein, ``Standardization
  and optimization of ct protocols to achieve low dose,'' {\em Journal of the
  American College of Radiology}, vol.~11, no.~3, pp.~271--278, 2014.

\bibitem{xue2012window}
Z.~Xue, S.~Antani, L.~R. Long, D.~Demner-Fushman, and G.~R. Thoma, ``Window
  classification of brain ct images in biomedical articles,'' in {\em AMIA
  Annual Symposium Proceedings}, vol.~2012, p.~1023, American Medical
  Informatics Association, 2012.

\bibitem{willemink2019evolution}
M.~J. Willemink and P.~B. No{\"e}l, ``The evolution of image reconstruction for
  ct—from filtered back projection to artificial intelligence,'' {\em
  European radiology}, vol.~29, no.~5, pp.~2185--2195, 2019.

\bibitem{yang2018low}
Q.~Yang, P.~Yan, Y.~Zhang, H.~Yu, Y.~Shi, X.~Mou, M.~K. Kalra, Y.~Zhang,
  L.~Sun, and G.~Wang, ``Low-dose ct image denoising using a generative
  adversarial network with wasserstein distance and perceptual loss,'' {\em
  IEEE transactions on medical imaging}, vol.~37, no.~6, pp.~1348--1357, 2018.

\bibitem{haque2020deep}
I.~R.~I. Haque and J.~Neubert, ``Deep learning approaches to biomedical image
  segmentation,'' {\em Informatics in Medicine Unlocked}, vol.~18, p.~100297,
  2020.

\bibitem{zhou2018performance}
X.~Zhou, K.~Yamada, T.~Kojima, R.~Takayama, S.~Wang, X.~Zhou, T.~Hara, and
  H.~Fujita, ``Performance evaluation of 2d and 3d deep learning approaches for
  automatic segmentation of multiple organs on ct images,'' in {\em Medical
  Imaging 2018: Computer-Aided Diagnosis}, vol.~10575, p.~105752C,
  International Society for Optics and Photonics, 2018.

\bibitem{weston2019automated}
A.~D. Weston, P.~Korfiatis, T.~L. Kline, K.~A. Philbrick, P.~Kostandy,
  T.~Sakinis, M.~Sugimoto, N.~Takahashi, and B.~J. Erickson, ``Automated
  abdominal segmentation of ct scans for body composition analysis using deep
  learning,'' {\em Radiology}, vol.~290, no.~3, pp.~669--679, 2019.

\bibitem{zhao2020mss}
W.~Zhao, D.~Jiang, J.~P. Queralta, and T.~Westerlund, ``Mss u-net: 3d
  segmentation of kidneys and tumors from ct images with a multi-scale
  supervised u-net,'' {\em Informatics in Medicine Unlocked}, p.~100357, 2020.

\bibitem{akbik2018contextual}
A.~Akbik, D.~Blythe, and R.~Vollgraf, ``Contextual string embeddings for
  sequence labeling,'' in {\em Proceedings of the 27th International Conference
  on Computational Linguistics}, pp.~1638--1649, 2018.

\bibitem{hassan2017deep}
A.~Hassan and A.~Mahmood, ``Deep learning for sentence classification,'' in
  {\em 2017 IEEE Long Island Systems, Applications and Technology Conference
  (LISAT)}, pp.~1--5, IEEE, 2017.

\bibitem{olah2016attention}
C.~Olah and S.~Carter, ``Attention and augmented recurrent neural networks,''
  {\em Distill}, vol.~1, no.~9, p.~e1, 2016.

\bibitem{badea2016use}
M.-S. Badea, I.-I. Felea, L.~M. Florea, and C.~Vertan, ``The use of deep
  learning in image segmentation, classification and detection,'' {\em arXiv
  preprint arXiv:1605.09612}, 2016.

\bibitem{lecun1998gradient}
Y.~LeCun, L.~Bottou, Y.~Bengio, and P.~Haffner, ``Gradient-based learning
  applied to document recognition,'' {\em Proceedings of the IEEE}, vol.~86,
  no.~11, pp.~2278--2324, 1998.

\bibitem{lin2013network}
M.~Lin, Q.~Chen, and S.~Yan, ``Network in network,'' {\em arXiv preprint
  arXiv:1312.4400}, 2013.

\bibitem{polsinelli2020light}
M.~Polsinelli, L.~Cinque, and G.~Placidi, ``A light cnn for detecting covid-19
  from ct scans of the chest,'' {\em arXiv preprint arXiv:2004.12837}, 2020.

\bibitem{okada2015abdominal}
T.~Okada, M.~G. Linguraru, M.~Hori, R.~M. Summers, N.~Tomiyama, and Y.~Sato,
  ``Abdominal multi-organ segmentation from ct images using conditional
  shape--location and unsupervised intensity priors,'' {\em Medical image
  analysis}, vol.~26, no.~1, pp.~1--18, 2015.

\bibitem{badrinarayanan2017segnet}
V.~Badrinarayanan, A.~Kendall, and R.~Cipolla, ``Segnet: A deep convolutional
  encoder-decoder architecture for image segmentation,'' {\em IEEE transactions
  on pattern analysis and machine intelligence}, vol.~39, no.~12,
  pp.~2481--2495, 2017.

\bibitem{ronneberger2015u}
O.~Ronneberger, P.~Fischer, and T.~Brox, ``U-net: Convolutional networks for
  biomedical image segmentation,'' in {\em International Conference on Medical
  image computing and computer-assisted intervention}, pp.~234--241, Springer,
  2015.

\bibitem{christ2017automatic}
P.~F. Christ, F.~Ettlinger, F.~Gr{\"u}n, M.~E.~A. Elshaera, J.~Lipkova,
  S.~Schlecht, F.~Ahmaddy, S.~Tatavarty, M.~Bickel, P.~Bilic, {\em et~al.},
  ``Automatic liver and tumor segmentation of ct and mri volumes using cascaded
  fully convolutional neural networks,'' {\em arXiv preprint arXiv:1702.05970},
  2017.

\bibitem{almotairi2020liver}
S.~Almotairi, G.~Kareem, M.~Aouf, B.~Almutairi, and M.~A.-M. Salem, ``Liver
  tumor segmentation in ct scans using modified segnet,'' {\em Sensors},
  vol.~20, no.~5, p.~1516, 2020.

\bibitem{zhou2016first}
X.~ZHOU, T.~ITO, R.~TAKAYAMA, S.~WANG, T.~HARA, and H.~FUJITA, ``First trial
  and evaluation of anatomical structure segmentations in 3d ct images based
  only on deep learning,'' {\em Medical Imaging and Information Sciences},
  vol.~33, no.~3, pp.~69--74, 2016.

\bibitem{long2015fully}
J.~Long, E.~Shelhamer, and T.~Darrell, ``Fully convolutional networks for
  semantic segmentation,'' in {\em Proceedings of the IEEE conference on
  computer vision and pattern recognition}, pp.~3431--3440, 2015.

\bibitem{de2015deep}
A.~de~Brebisson and G.~Montana, ``Deep neural networks for anatomical brain
  segmentation,'' in {\em Proceedings of the IEEE conference on computer vision
  and pattern recognition workshops}, pp.~20--28, 2015.

\bibitem{roth2015deep}
H.~R. Roth, A.~Farag, L.~Lu, E.~B. Turkbey, and R.~M. Summers, ``Deep
  convolutional networks for pancreas segmentation in ct imaging,'' in {\em
  Medical Imaging 2015: Image Processing}, vol.~9413, p.~94131G, International
  Society for Optics and Photonics, 2015.

\bibitem{cha2016urinary}
K.~H. Cha, L.~Hadjiiski, R.~K. Samala, H.-P. Chan, E.~M. Caoili, and R.~H.
  Cohan, ``Urinary bladder segmentation in ct urography using deep-learning
  convolutional neural network and level sets,'' {\em Medical physics},
  vol.~43, no.~4, pp.~1882--1896, 2016.

\bibitem{zhou2016three}
X.~Zhou, T.~Ito, R.~Takayama, S.~Wang, T.~Hara, and H.~Fujita,
  ``Three-dimensional ct image segmentation by combining 2d fully convolutional
  network with 3d majority voting,'' in {\em Deep Learning and Data Labeling
  for Medical Applications}, pp.~111--120, Springer, 2016.

\bibitem{kayid2018performance}
A.~Kayid, Y.~Khaled, and M.~Elmahdy, ``Performance of cpus/gpus for deep
  learning workloads,'' {\em The German University in Cairo}, 2018.

\bibitem{cciccek20163d}
{\"O}.~{\c{C}}i{\c{c}}ek, A.~Abdulkadir, S.~S. Lienkamp, T.~Brox, and
  O.~Ronneberger, ``3d u-net: learning dense volumetric segmentation from
  sparse annotation,'' in {\em International conference on medical image
  computing and computer-assisted intervention}, pp.~424--432, Springer, 2016.

\bibitem{milletari2016v}
F.~Milletari, N.~Navab, and S.-A. Ahmadi, ``V-net: Fully convolutional neural
  networks for volumetric medical image segmentation,'' in {\em 2016 fourth
  international conference on 3D vision (3DV)}, pp.~565--571, IEEE, 2016.

\bibitem{chen2016voxresnet}
H.~Chen, Q.~Dou, L.~Yu, and P.-A. Heng, ``Voxresnet: Deep voxelwise residual
  networks for volumetric brain segmentation,'' {\em arXiv preprint
  arXiv:1608.05895}, 2016.

\bibitem{alalwan60efficient}
N.~Alalwan, A.~Abozeid, A.~A. ElHabshy, and A.~Alzahrani, ``Efficient 3d deep
  learning model for medical image semantic segmentation,'' {\em Alexandria
  Engineering Journal}, vol.~60, no.~1, pp.~1231--1239.

\bibitem{li2018h}
X.~Li, H.~Chen, X.~Qi, Q.~Dou, C.-W. Fu, and P.-A. Heng, ``H-denseunet: hybrid
  densely connected unet for liver and tumor segmentation from ct volumes,''
  {\em IEEE transactions on medical imaging}, vol.~37, no.~12, pp.~2663--2674,
  2018.

\bibitem{simonyan2014very}
K.~Simonyan and A.~Zisserman, ``Very deep convolutional networks for
  large-scale image recognition,'' {\em arXiv preprint arXiv:1409.1556}, 2014.

\bibitem{chung2020ct}
M.~Chung, A.~Bernheim, X.~Mei, N.~Zhang, M.~Huang, X.~Zeng, J.~Cui, W.~Xu,
  Y.~Yang, Z.~A. Fayad, {\em et~al.}, ``Ct imaging features of 2019 novel
  coronavirus (2019-ncov),'' {\em Radiology}, vol.~295, no.~1, pp.~202--207,
  2020.

\bibitem{yushkevich2016itk}
P.~A. Yushkevich, Y.~Gao, and G.~Gerig, ``Itk-snap: An interactive tool for
  semi-automatic segmentation of multi-modality biomedical images,'' in {\em
  2016 38th Annual International Conference of the IEEE Engineering in Medicine
  and Biology Society (EMBC)}, pp.~3342--3345, IEEE, 2016.

\bibitem{lee2018practical}
H.~Lee, M.~Kim, and S.~Do, ``Practical window setting optimization for medical
  image deep learning,'' {\em arXiv preprint arXiv:1812.00572}, 2018.

\bibitem{woo2018cbam}
S.~Woo, J.~Park, J.-Y. Lee, and I.~So~Kweon, ``Cbam: Convolutional block
  attention module,'' in {\em Proceedings of the European conference on
  computer vision (ECCV)}, pp.~3--19, 2018.

\bibitem{hesamian2019deep}
M.~H. Hesamian, W.~Jia, X.~He, and P.~Kennedy, ``Deep learning techniques for
  medical image segmentation: Achievements and challenges,'' {\em Journal of
  digital imaging}, vol.~32, no.~4, pp.~582--596, 2019.

\bibitem{chollet2017xception}
F.~Chollet, ``Xception: Deep learning with depthwise separable convolutions,''
  in {\em Proceedings of the IEEE conference on computer vision and pattern
  recognition}, pp.~1251--1258, 2017.

\bibitem{hahn2016comparison}
K.~Hahn, H.~Sch{\"o}ndube, K.~Stierstorfer, J.~Hornegger, and F.~Noo, ``A
  comparison of linear interpolation models for iterative ct reconstruction,''
  {\em Medical physics}, vol.~43, no.~12, pp.~6455--6473, 2016.

\end{thebibliography}
\end{document}